          [2001/04/25 1.1 (PWD)]
\documentclass[a4paper,twocolumn]{esapub} 
\usepackage{natbib}
\usepackage{epsfig}

\title{INTEGRAL observations of the Large Magellanic Cloud region}
\author{S.Mereghetti}
\author{D.G\"{o}tz}
\author{A.Paizis}
\author{A.Pellizzoni}
\author{S.Vercellone}
\affil{ Istituto di Astrofisica Spaziale e Fisica Cosmica,
Sezione di Milano  G.Occhialini, CNR, Italy}
\author{N.J.Westergaard}
\affil{Danish Space Research Institute Copenhagen, Denmark}
\author{O.Vilhu}
\affil{ Observatory University of Helsinki, Finland}
\author{T.Belloni}
\affil{INAF - Osserv. Astronomico di Brera, Italy}
\author{R.Walter}
\author{T.Courvoisier}
\author{K.Ebisawa}
\author{P.Kretschmar}
\affil{INTEGRAL Science Data Center, Versoix, Switzerland}
\author{L.Stella}
\affil{INAF - Osserv. Astronomico di Roma, Monteporzio, Italy}
\author{J.-P.Swings}
\affil{Institut d'Astrophysique et de Geophysique, Liege, Belgium}
\author{J.Kn\"{o}dlseder}
\affil{CESR, Toulouse, France}
\author{A.Dean}
\affil{Southampton University, United Kingdom}
\author{A.Strong}
\affil{MPE, Garching, Germany}
\author{P.Hakala}
\affil{Tuorla Observatory,  Turku, Finland}
\author{A.Zdziarski}
\affil{N.Copernicus Astronomical Ctr., Warsaw, Poland}

\begin{document}


\maketitle

\begin{abstract}
We present the preliminary results of the INTEGRAL survey of the
Large Magellanic Cloud. The observations have been carried out in
January 2003 (about 10$^6$ s) and January 2004 (about
4$\times10^{5}$ s). Here we concentrate on the bright sources LMC
X-1, LMC X-2, LMC X-3 located in  our satellite galaxy, and on the
serendipitous detections of the Galactic Low Mass X-ray Binary EXO
0748--676 and of the Seyfert 2 galaxy IRAS 04575--7537.
\end{abstract}

\section{Introduction}

\begin{figure*}[ht!]
\centerline{\psfig{figure=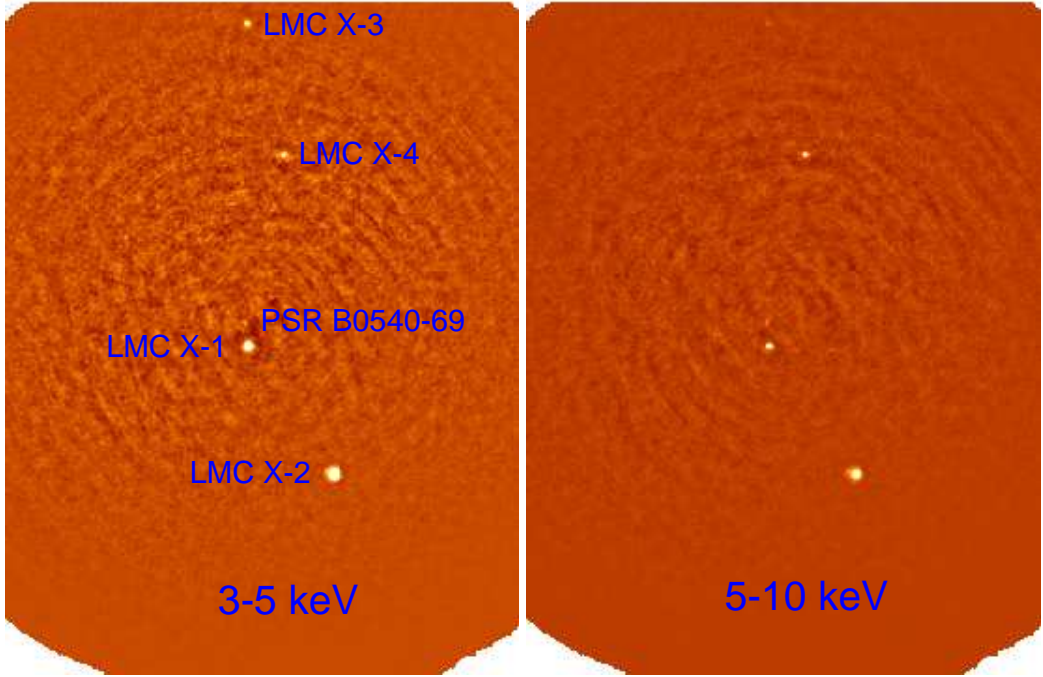,angle=0,width=14cm}}
\caption{Mosaics of the JEM-X observations of the LMC field
performed in January 2003.} \label{jemx-im}
\end{figure*}

The Large Magellanic Cloud (LMC), with an angular extent of
$\sim$10$^{\circ}$$\times$5$^{\circ}$ well matched to the fields
of view of the IBIS \citep{ibis} and SPI \citep{spi} instruments,
is an ideal target to exploit the multiplexing capabilities of
INTEGRAL.

Being the closest galaxy (d$\sim$54 kpc), the LMC has always been
a primary target for all  X-ray and $\gamma$-ray missions. It
contains a few  luminous  X-ray binaries, that have been
extensively studied in X-rays,  but for which little information
has been obtained above $\sim$20-30 keV. Besides giving
information on the brightest sources, an INTEGRAL survey of the
LMC  should allow studies of diffuse emission, and to discover new
fainter hard X-ray sources.

INTEGRAL observed the LMC for about 1 Msec in January 2003 and for
about 400 ksec in January 2004. Fig.\ref{jemx-im} shows the images
of the LMC field obtained with the JEM-X instrument \citep{jemx}
in the energy ranges 3-5 keV and 5-10 keV.

Here we present the preliminary results on the brightest sources
detected in this field with IBIS: the black hole candidate LMC
X-1,  the Low Mass X-ray Binary EXO 0748--676, and the AGN IRAS
04575--7537. The two latter sources do not belong to the LMC, but
were serendipitously detected in the large IBIS field of view.  As
can be seen in Fig.\ref{jemx-im}, also LMC X-2, LMC X-3 and LMC
X-4\footnote{we will not report on LMC X-4 since the data rights
for this source belong to a different group} were clearly
detected. LMC X-2 is a Low Mass X-ray Binary containing a neutron
star, and the only Z-type source known out of our Galaxy
\citep{smale}. LMC X-3 is a persistent black hole candidate
\citep{cowley} which spends most of the time in a soft spectral
state, with a very faint power law tail at high energies. Due to
their  soft spectrum and relatively low flux, LMC X-2 and LMC X-3
were not detected in the preliminary analysis of the IBIS data
carried out to date.

\section{LMC X--1}

LMC X-1 is a persistent black hole candidate (BHC) with an
accurately measured mass function (0.14 $M_{sun}$, \citet{hutch}).
The most likely mass for the compact object is $\sim$4 $M_{sun}$.
To date, this BHC has always been found  in a typical high/soft
state with  luminosity  L$_X$$\sim$2 10$^{38}$ erg s$^{-1}$ and a
spectrum well described by a disk-blackbody component with rather
constant temperature $\sim$0.9 keV plus a variable, steep
power-law tail with photon index $\Gamma\sim$2.5--3
\citep{ebisawa,haardt}.

\begin{figure}[ht!]
\centerline{\psfig{figure=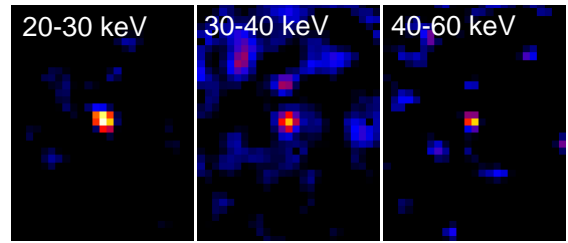,angle=0,width=7.5cm}}
\caption{Images of the region of LMC X--1 obtained with ISGRI in
different energy ranges. Each image has a size of
$\sim5^{\circ}\times5^{\circ}$} \label{lmcx1-im}
\end{figure}

Previous high-energy observations detected LMC X-1 up to $\sim$60
keV, but they were carried out with non-imaging instruments and
might have suffered from confusion problems, in particular due to
the nearby radio pulsar PSR 0540--69, which lies at an angular
distance of $\sim$25 arcmin. With ISGRI we obtained the first
images of LMC X-1 above 20 keV (see Fig.\ref{lmcx1-im}).

\begin{figure}[ht!]
\centerline{\psfig{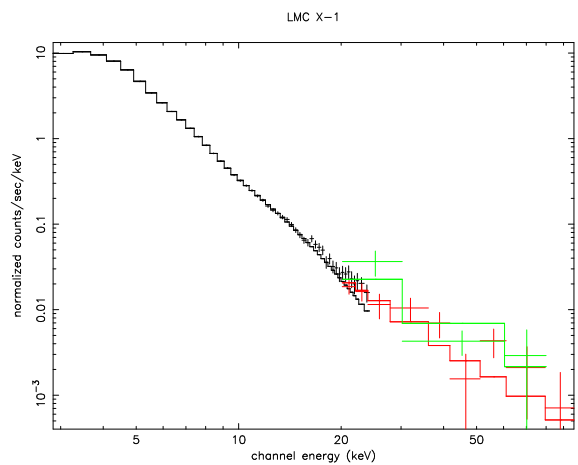}}
\caption{Combined JEM-X, IBIS/ISGRI and SPI energy spectrum of LMC
X-1. The best fit parameters are kT$_{in}$=1.9$\pm$0.5 keV and
photon index 2.7$\pm$0.4} \label{lmcx1-sp}
\end{figure}

We performed a joint spectral analysis of the 2003 data from ISGRI
and JEM-X. The resulting spectrum is shown in Fig.\ref{lmcx1-sp}.
The best fit parameters for a model composed by a disk blackbody
\citep{mitsuda} plus a power-law (kT$_{in}$=1.9$\pm$0.5 keV,
photon index = 2.7$\pm$0.4) are quite different from those found
in previous observations of this source. This might simply reflect
systematics in the low energy calibration currently available for
JEM-X.

\section{EXO 0748--676}

\begin{figure*}[ht!]
\centerline{\psfig{figure=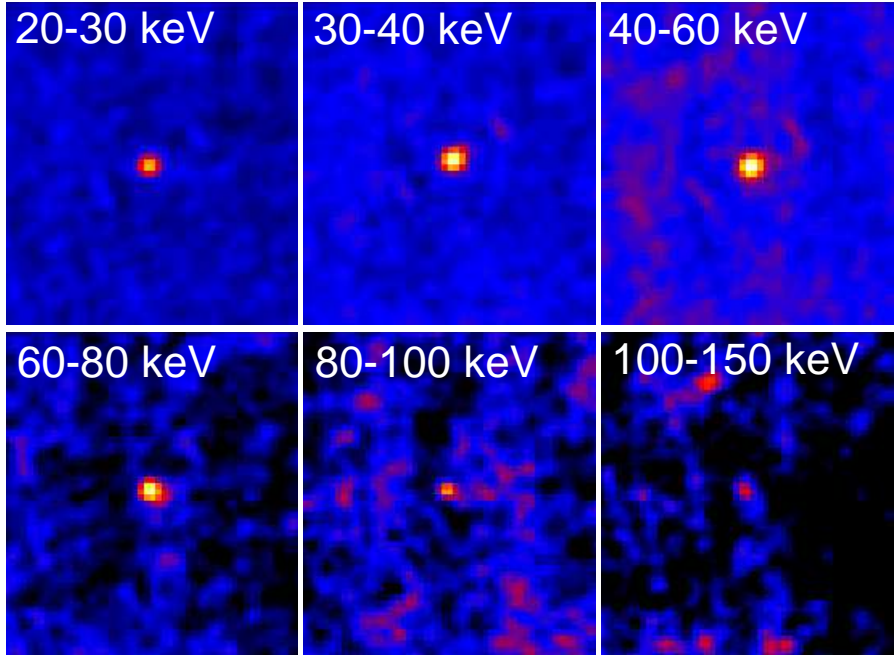,angle=0,width=12cm}}
\caption{ISGRI images of EXO 0748--676 in different energy ranges.
Each image has a size of $\sim5^{\circ}\times5^{\circ}$.}
\label{exo-im}
\end{figure*}

\begin{figure}[ht!]
\centerline{\psfig{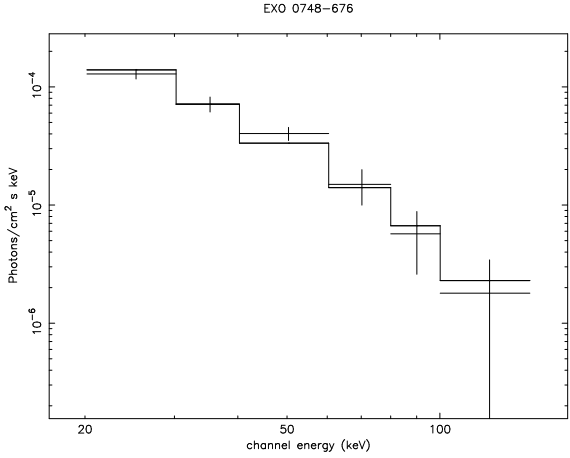}}
\caption{IBIS/ISGRI energy spectrum  of EXO 0748--676.}
\label{exo-sp}
\end{figure}

EXO 0784--676 is a dipping Low Mass X-ray Binary containing a
neutron star. It has been extensively studied at low energies
($<$10 keV), where it shows  all types of variability typical of
the different kinds of LMXRB: eclipses, dips, flares and Type I
bursts \citep{homan}. This source is not located in the LMC, but
falls serendipitously in the IBIS field of view of our
observations, and, despite its rather large off-axis angle it has
been detected in all the energy bands of our ISGRI mosaic (see
Fig.\ref{exo-im}).

Fig.\ref{exo-sp} shows its spectrum as measured with ISGRI. It is
well fit by a cut-off power law with photon index $\sim$1.4 and
$E_c$=50 keV.  The time averaged flux in the 20-100 keV range was
2$\times10^{-10}$ erg cm$^{-2}$ s$^{-1}$, about 40\% lower than
that measured by the BeppoSAX PDS instrument in November 2000
\citep{sidoli}.

\section{IRAS 04575--7537}

\begin{figure}[ht!]
\centerline{\psfig{figure=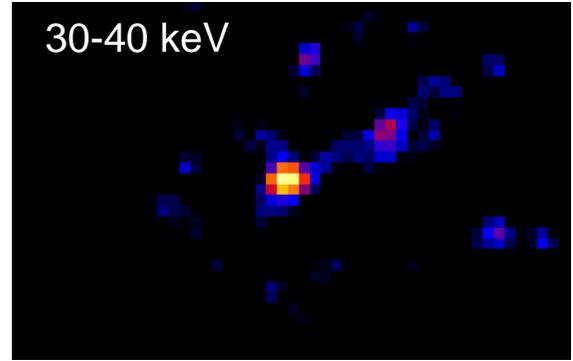,angle=0,width=7.5cm}}
\caption{ISGRI image of the Seyfert 2 galaxy IRAS 04575--7537.}
\label{iras-sp}
\end{figure}

\begin{figure}[ht!]
\centerline{\psfig{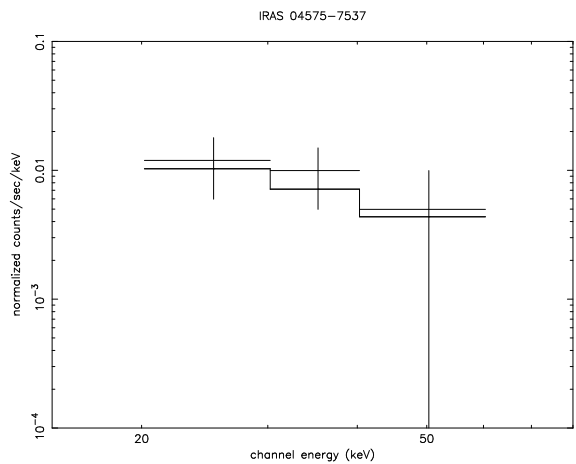}}
\caption{ISGRI energy spectrum of  IRAS 04575--7537. It can be fit
by a power law with photon index $\sim$1.5.} \label{iras-sp}
\end{figure}

IRAS 04575--7537 is a background AGN falling in the LMC field. It
is associated with a rather bright barred spiral Galaxy at
redshift z=0.0184 and showing a  strong infrared emission
\citep{hew}. IRAS 04575--7537 was first detected in the X-ray
range with HEAO-1, and later observed with ROSAT, GINGA and ASCA
\citep{vignali}. It is classified as a Seyfert 2 galaxy.

In our observations it was quite faint ($\sim$3 mCrab) and
detected with a signal to noise of 7 in the two lower energy bands
of the ISGRI mosaic.  We have derived two spectral points and an
upper limit, which represent the first measurement of this AGN at
these energies. The flux is about   1.1$\times10^{-11}$ erg
cm$^{-2}$ s$^{-1}$ (20-40 keV). Extrapolating the oserved spectrum
to higher energies would imply a 20-100 keV luminosity of
5$\times10^{43}$ erg s$^{-1}$, comparable to that in the far
infrared band \citep{ward}.

\section*{Acknowledgments}

This work has been partly funded by ASI.

\end{document}